\documentclass[twocolumn,showpacs,preprintnumbers,amsmath,amssymb]{revtex4}

\usepackage{graphicx}
\usepackage{dcolumn}
\usepackage{bm}
\usepackage[usenames]{color}
\usepackage{ulem}

\begin{document}


\title{Size Dependence of Current Spin Polarization Through Superconductor/Ferromagnet Nanocontacts
}

\author{M. Stokmaier$^1$}
\author{G. Goll$^1$}
\email{gernot.goll@phys.uni-karlsruhe.de}
\author{D. Weissenberger$^{2,3}$}
\author{C. S\"urgers$^{1,2}$}
\author{H. v. L\"ohneysen$^{1,2,4}$}
\email{h.vl@phys.uni-karlsruhe.de}
\affiliation{%
$^1$Physikalisches Institut and $^2$DFG Center for Functional Nanostructures, Universit\"at Karlsruhe, 76128 Karlsruhe, Germany}
\affiliation{%
$^3$Laboratorium f\"ur Elektronenmikroskopie, Universit\"at Karlsruhe, 76128 Karlsruhe, Germany}
\affiliation{$^4$Forschungszentrum Karlsruhe, Institut f\"ur Festk\"orperphysik, 76021 Karlsruhe, Germany}
\date{\today}

\begin{abstract}
The spin polarization $P$ of the transport current through the interface between superconducting Al and ferromagnetic Fe is determined by means of Andreev reflection at nanostructured point contacts. We observe a systematic decrease of $P$ with decreasing contact resistance. Our data provide evidence for the reduction of $P$ by spin-orbit scattering and thus establish a link between density-of-states and transport spin polarizations.
\end{abstract}

\maketitle

The spin-dependent electronic transport in magnetic heterostructures is an emergent field because of potential applications in spintronics \cite{zut04,par04}. The knowledge of the current spin polarization $P$ of materials is a key issue for the functionality \cite{pri95}. This is particularly important for high-$P$ materials such as half-metallic alloys, e.\,g., Heusler compounds \cite{gal05,wur06}. The spin polarization of ferromagnets can be measured by various techniques, including spin-polarized photoemission \cite{joh97}, spin-dependent electron tunneling \cite{mes94}, and point-contact Andreev reflection (AR) between a ferromagnet F and a superconductor S \cite{jon95}. An issue of considerable importance is how the current spin polarization obtained by AR \cite{and64} is related to the ferromagnet's bulk spin polarization \cite{xia02}.

AR has been widely used to extract the current spin polarization in a variety of materials \cite{sou98,upa98,nad00,ji01,ray03,par02}. In particular, Upadhyay \textit{et al.} \cite{upa98} employed the type of contacts also used by us. AR at an interface between a normal metal N and S is the coherent process by which an electron from N incident on S is retro-reflected as a hole, thus creating a Cooper pair in S. The theoretical analysis of these experiments is usually carried out in the spirit of
the Blonder-Tinkham-Klapwijk (BTK) theory \cite{blo82}. The ordinary reflection at the N/S interface is parametrized by a phenomenological parameter $Z$. The sensitivity of the Andreev process to the spin of the carriers originates from the spin content of a Cooper pair ($S=0$ in conventional superconductors) and the conservation of the spin direction at the interface. Employing a ferromagnet as the nonsuperconducting electrode leads to a reduction of the AR probability for spin-singlet superconductors because incoming electron and outgoing hole must have opposite spins \cite{jon95}. Two major approaches have been suggested to extract $P$ from AR datas: (i) the AR current is decomposed into a fully spin-polarized and fully unpolarized current \cite{xia02,maz01,woo04}, (ii) spin-dependent transmission coefficients $\tau_{\uparrow}$ and $\tau_{\downarrow}$ are introduced, hence $P = \mid \tau_{\uparrow} - \tau_{\downarrow}\mid / (\tau_{\uparrow} + \tau_{\downarrow})$ \cite{per04}. For a comprehensive recent review of different models to extract the AR spin polarization, see Ref.\,\onlinecite{cha07}. In this letter, we report on AR spectra obtained on a series of nanostructured Al/Fe contacts that show a distinct dependence on contact size. A possible origin of this systematic size dependence is spin-orbit scattering, shedding light on the origin of the differences in the spin polarization determined by different experimental methods. We chose Fe in this work because it does not show surface degradation of $P$ when prepared under UHV, in contrast to many Heusler compounds that show a reduced $P$ at the surface.

We employ nanostructured Al/Fe point contacts (PCs) that are formed by two metal films connected by  a tiny hole in an insulating membrane in between \cite{ral89}. These contacts are very stable and reproducible spectra are obtained even after several thermal cycles. Furthermore, \textit{in-situ} preparation of the metal films in a single run avoids oxidation or degradation of the interface between the two electrodes. Holes of radius $a$ between 7.5 and 25\,nm were fabricated by e-beam structuring of a 50-nm thick Si$_3$N$_4$. $a$ was determined from the contact resistance (see below) after evaporation in UHV of a 200-nm thick Al layer on the side of the membrane with a bowl-like funnel towards the hole, and of a 12-nm thick Fe layer topped by a Cu layer of thickness $d_{\rm Cu} = 188$\,nm on the opposite (flat) side (see Refs.\,\onlinecite{ral89,per04} for details). By variation of the process parameters for the e-beam lithographical prestructuring and subsequent reactive ion-etching,  contacts with resistance $R_N$ between $1\,\Omega$ and $\approx 30\,\Omega$ are obtained. Low-ohmic PCs ($R_N< 2\,\Omega$) implying a large contact diameter usually show clear evidence of non-ballistic transport and therefore were not used for the present investigation. The differential conductance $G={\rm d}I/{\rm d}V$ vs. applied bias voltage $V$ was measured in a dilution refrigerator between $T\approx 20$\,mK and 2\,K.

\begin{figure}
\includegraphics[width=60mm]{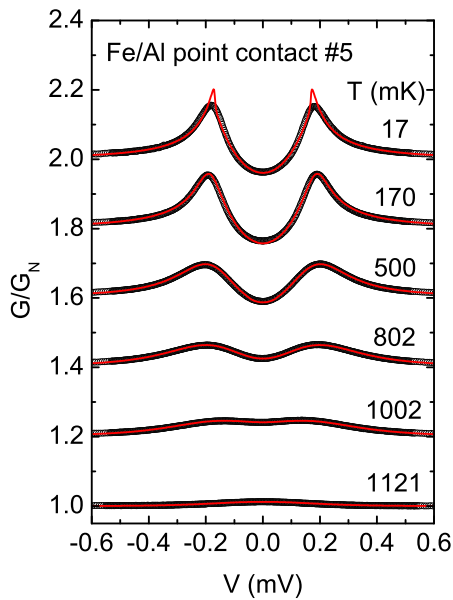}
\caption{\label{fig:stok_f1} (Color online) Differential conductance 
$G={\rm d}I/{\rm d}V$ normalized to the normal-state value $G_{\rm N}=G(T>T_c) = 54.3\,{\rm mS}$ for a nanostructured Al/Fe point contact (sample \#5) for different $T$. For clarity, the curves are shifted upwards successively by 0.2 units with decreasing $T$. The conductance curves calculated with the $\tau_{\uparrow}$- $\tau_{\downarrow}$ model match the experimental curves exactly except for the lowest $T$ in the region of the $G$ maximum (solid line).}
\end{figure}

\begin{figure}
\includegraphics[width=55mm]{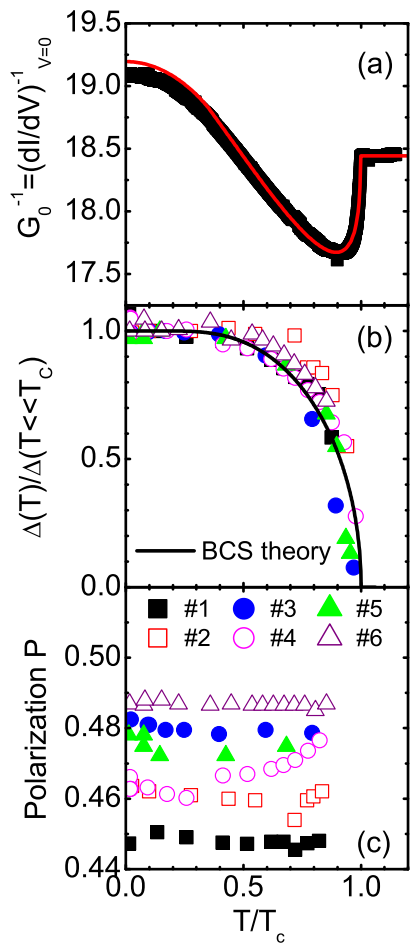}
\caption{\label{fig:stok_f3} (Color online) (a): $T$ dependence of the zero-bias resistance $G_0^{-1}$ together with a theoretical curve (see text). (b): $T$ dependence of the superconducting gap $\Delta$ for all samples investigated in this study, obtained from a least-squares fit of the data and normalized to $\Delta (T<200\,\mathrm{mK}\ll T_c)$. (c): $T$ dependence of the current spin polarization $P=\mid \tau_{\uparrow} - \tau_{\downarrow}\mid / (\tau_{\uparrow} + \tau_{\downarrow})$. $T$ is normalized to the experimental $T_c^{\mathrm{exp}}$ of each sample in order to account for the small variation of $T_c$ from sample to sample.}
\end{figure}

Fig.\,\ref{fig:stok_f1} exemplarily shows the differential conductance $G(V)$ of a PC with $R_N(4.2\,{\rm K})=18.4\,\Omega$ at various temperatures $T$ below $T_c$ of the Al film. $G$ is normalized to the normal-state conductance $G_N= R_N^{-1}$. The data are analyzed in terms of two spin-dependent transmission coefficients $\tau_{\uparrow}$ and $\tau_{\downarrow}$ for the majority and minority charge carriers in F, respectively. We assume that the PC can be described with a single pair of transmission coefficients \cite{mar01,per04} which contain the microscopic properties relevant for the transport, i.\,e., the spin-split band structure of F, the electronic structure of S, and the interface properties.
For $\tau < 1$, an electron experiences ordinary reflection with a finite probability $1- \tau$. This leads to a minimum of $G$ at zero bias and sharp maxima at voltages corresponding to the superconducting energy gap $\Delta (T)$. Using $\tau_{\uparrow}$, $\tau_{\downarrow}$, and $\Delta$ as free parameters for each $T$, the model yields excellent least-squares fits to the Andreev spectra over the whole $T$ range below $T_c$ and the entire voltage range, as shown in Fig.\,\ref{fig:stok_f1}. Only at the lowest nominal $T=17$\,mK, where the fit yields $\tau_{\uparrow} = 0.348$, $\tau_{\downarrow} = 0.997$, and $\Delta = 166 \,\mu$eV, can the fits be distinguished from the data, the calculated maxima appear somewhat sharper than experimentally observed. This might be caused by a levelling-off of the electron $T$ due to heating by electromagnetic stray fields or a small pair-breaking effect by Fe. The excellent agreement between experiment and theory is also reflected in the $T$ dependence of the zero-bias resistance $G_0^{-1}=({\rm d}I/{\rm d}V)^{-1}_{V=0}$ shown in Fig.\,\ref{fig:stok_f3}a. $P$ is found to be 0.482 for this particular sample. The characteristic parameters for a total of six contacts are listed in Table\,\ref{tab:table1}.
\begin{figure}
\includegraphics[width=60mm,clip=]{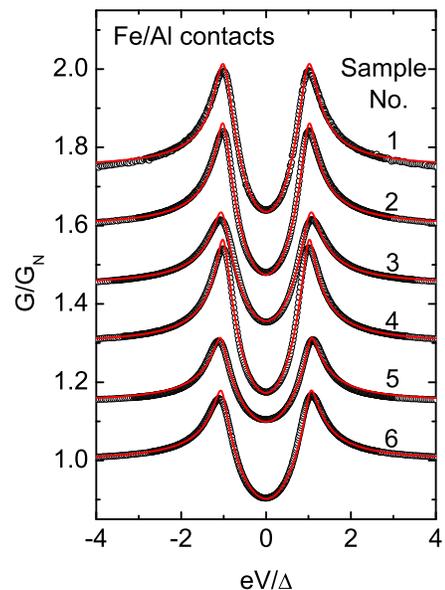}
\caption{\label{fig:stok_f2} (Color online) Normalized differential conductance 
$G/G_{\rm N}$ of nanostructured Al/Fe point contacts at $T\approx 0.1$\,K. For clarity, the curves are shifted upwards successively by 0.15 units. The conductance curves (solid lines) are calculated with the $\tau_{\uparrow}$- $\tau_{\downarrow}$ model (see text).}
\end{figure}

Shown in Fig.\,\ref{fig:stok_f2} are $G(V)$ curves obtained at $T\approx 0.1$\,K for all samples. One clearly sees a systematic reduction of the amplitude of the AR-related features when going from sample 1 (small $R_N$) to sample 6 (large $R_N$), indicating a systematic increase of $P$.Again, the quality of the fits is excellent over the whole voltage range up to $V=4\Delta_0/e$.

\begin{table}
\caption{\label{tab:table1} Superconductive energy gap $\Delta_0=\Delta(T\rightarrow 0)$, transition temperature $T_c^{\mathrm{exp}}$, transmission coefficients $\tau_\uparrow$, $\tau_\downarrow$, and current spin polarization $P$ as determined by a fit of the Andreev spectra at $T<200$\,mK for point contacts with contact resistance $R_N=1/G_{\rm N}$ at $T=4.2$\,K. $a$ is the contact radius calculated from the Wexler resistance (see text).}
\begin{ruledtabular}
\begin{tabular}{cccccccc}
Sample& $R_N$ &$a$& $\Delta_0$ & $T_c^{\mathrm{exp}}$ & $\tau_\uparrow$ &$\tau_\downarrow$ & $P$ \\
No.&$(\Omega)$&(nm)&(meV)&(mK)&&&\\
\hline
1 & 2.68 &24.7& 0.174 &971& 0.371 & 0.983 & 0.452 \\
2 & 6.98 &14.3& 0.175 &903& 0.362 & 0.984 & 0.460 \\
3 & 7.29 &13.9& 0.157 &1009& 0.349 & 0.993 & 0.480 \\
4 & 9.59 &12.0& 0.190 &970& 0.361 & 0.984 & 0.462 \\
5 & 18.4 &8.4& 0.166 &1174& 0.348 & 0.997 & 0.482 \\
6 & 24.2 &7.3& 0.174 &1115& 0.343 & 0.994 & 0.487 \\
\end{tabular}
\end{ruledtabular}
\end{table}

The high stability of our PCs, together with the high quality of the fits, allows a detailed investigation of the $T$ dependence of $\Delta (T)$ and $P(T)$ as  displayed in Fig.\,\ref{fig:stok_f3}b and \ref{fig:stok_f3}c. $T$ is normalized to the experimental $T_c^{\mathrm{exp}}$ of each sample in order to account for the minor variation of $T_c$ from sample to sample. $T_c^{\mathrm{exp}}$ has been determined from the sudden drop of the zero-bias resistance $G^{-1}_0(T)$ (see Fig.\,\ref{fig:stok_f3}a). $\Delta(T)$ follows the standard BCS dependence (solid line in  Fig.\,\ref{fig:stok_f3}b).
At first sight, this might be surprising because the superconductor is in direct contact with a ferromagnet. Apparently, pair breaking plays a minor role as also inferred from the small rounding of the AR spectra mentioned above. Here, the small dimensions of the S/F contact are instrumental, because they limit the AR laterally to a tiny region and thus do not strongly affect S. Indeed, $T_c$ appears to be somewhat lower for low $R_{\rm N}$, i.\,e., larger contacts (see Table\,\ref{tab:table1}). The current spin polarization $P$ for each sample is independent of $T$ within the error of the least-squares fitting procedure (Fig.\,\ref{fig:stok_f3}c). Only close to $T_c$ where the spectra become very shallow does the scatter in $P$ increase.  The finding that $P$ is independent of $T$ strongly supports our assignment of $P$ as an intrinsic parameter of the particular S/F PC under study.

Inspection of Table\,1 reveals that indeed $P$ depends on $R_N$ in a systematic fashion, i.e., larger contact resistances go along with larger $P$. We interpret this finding as follows. In view of the fact that all samples were prepared in the same fashion, we are confident that it is chiefly the contact radius $a$ that determines $R_N$. In particular, we recall that the interface between Al and Fe was never exposed to air. The radius $a$ of a circular homocontact can be calculated from $R_N$ for different regimes of current flow through the contact, depending on the ratio of $a$ to the electron mean free path $\ell$ of the material, with the limiting cases of the ballistic or Sharvin regime if $a\ll \ell$ and the thermal or Maxwell regime if $a\gg \ell$. For the intermediate regime an interpolation for $R_N=\gamma (K)\rho /(2a)+4\rho K/(3\pi a)$ was suggested by Wexler \cite{wex66}. Here $\gamma (K)$ is a slowly varying function of $K=\ell /a$ with $\gamma (0)=1$ and $\gamma (K\rightarrow \infty)=0.694$ and the Maxwell ($K\rightarrow 0$) and Sharvin limits ($K\rightarrow \infty$) are given by $R_N^{K\rightarrow 0}=\rho /2a$ and $R_N^{K\rightarrow \infty}=4\rho \ell/3\pi a^2$ \cite{sha65}, respectively. Using arithmetically averaged parameters $\rho \ell = 2.66\cdot 10^{-15}\,\Omega\mathrm{m}^2$ and $\rho=5.84\cdot 10^{-8}\,\Omega$m \cite{rem1}, and solving the Wexler formula iteratively for self-consistent values of $K$, $a$ is found to vary between 7.3 and 24.7\,nm, i.e., the change in $R_N$ corresponds to a change in $a$ by a factor of 3 (see Table\,1).

\begin{figure}
\includegraphics[width=75mm]{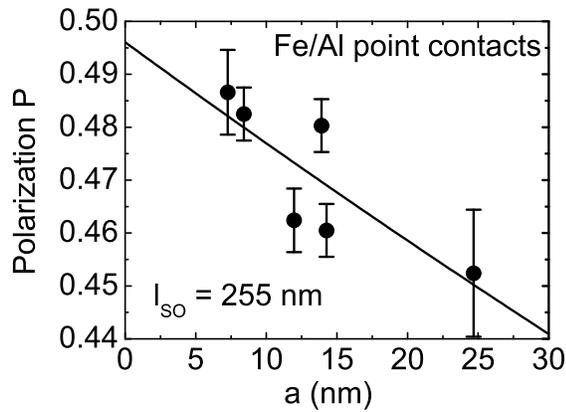}
\caption{\label{fig:stok_f4} Dependence of the current spin polarization on the contact radius. The solid line shows $P = P_0\exp{(-a/\ell_{so})}$ with $P_0 = 0.496\pm 0.01$ and $\ell_{so}=255\pm 91$\,nm. Over the short distance $a\ll \ell_{so}$ the exponential decay of $P(a)$ looks almost like a straight line.}
\end{figure}

$P(a)$ decreases systematically as displayed in Fig.\,\ref{fig:stok_f4}. The error bars reflect the statistical error of typically 6-8 measurements, all taken below 200\,mK $\approx 0.2\,T_c$ for each sample. We note that the overall trend of $P(a)$ does not depend on the details of how $a$ was calculated. A possible explanation of the $P(a)$ dependence is as follows. For  large $a$ the incident electrons  probe a larger volume before being retro-reflected as a hole than for a small $a$. Any scattering process affecting $P$ will therefore have a stronger effect for larger contacts \cite{nai04}. We suggest that spin-orbit scattering $-$ with a constant scattering length $-$ is operative as the main source of the reduction of $P$, and model the dependence $P(a)$ by a simple exponential decay, $P(a) = P_0\exp{(-a/\ell_{so})}$. We obtain a spin-orbit scattering length $\ell_{so}  = 255\pm 91$\,nm from a fit to the data in Fig.\,\ref{fig:stok_f4}. This value is very reasonable albeit somewhat lower than the spin diffusion length $\lambda_\mathrm{s}^\mathrm{N}=500$ - 1000\,nm found for Al \cite{bec04}, which -- apart from the error introduced by the scatter of the data -- may be attributed to a smaller elastic mean free path of our Al films or minor interdiffusion at the Al/Fe interface \cite{bra06}. We note that using the model of Mazin \textit{et al.} \cite{maz01} with the fit parameters $\Delta$, $P$, and $Z$, the same systematic trend of $P(a)$ with even smaller $\ell_{so}$ is obtained with no systematic correlation between $P$ and $Z$.
With this interpretation, our data constitute the first instance of a clear relation between the spin polarization of the Andreev current and the bulk spin polarization. $P_0=0.496\pm 0.01$ almost reaches the value obtained by spin-polarized photoemission spectroscopy on polycrystalline Fe films \cite{bus71} and on the $(111)$ surface of single-crystalline Fe \cite{eib78} yielding $P=0.54$ and $P=0.6$, respectively. The domain structure of the Fe film is expected to play no appreciable role for our PCs because both the domain size and the thickness of domain walls are much larger than the contact size \cite{hub98}.

For Pb/Co and Pb/Ni nanostructured PCs the current polarization was reported to be independent of the contact size with $\Delta G/G_{\rm N}=\pm 0.2$ at any (fixed) voltage \cite{upa98}. We note that our systematic variation of $\Delta G/G_{\rm N}=\pm 0.2$ at any voltage is also of that order. However, given the fact that Pb exhibits larger spin-orbit scattering, a larger variation would be expected. Possibly, the different behavior originates from a different microscopic structure of the interface and/or disorder in the PC region, perhaps from interfacial spin-dependent scattering whose importance was demonstrated by investigations of the influence of the Co film thickness on the spin-filtering effect in nanostructered Pb/Co contacts \cite{upa99}. Indeed $P$ obtained for the Pb/Co and Pb/Ni contacts is substantially lower than $P$ of our Al/Co \cite{per04} and Al/Fe contacts. This underscores the need to prepare samples under identical conditions in order to determine the dependence of the current spin polarization on the contact radius as done in the present work.

In conclusion, we have found a systematic dependence of the spin polarization $P$ of the Andreev current in S/F contacts on the contact size. We suggest that this size dependence arises from spin-orbit scattering which reduces $P$ and is most effective in large contacts. This size effect must be taken into account when determining $P$ via Andreev reflection.

We thank J. C. Cuevas, W. Belzig, and K. Samwer for helpful discussions.


\begin{thebibliography}{10}
\bibitem{rem1} The contact resistance of a heterocontact is naturally defined as the sum of the contributions of both metals \cite{nai04}, i.\,e. for the resistivity $\rho=(\rho_{\mathrm{Al}}+\rho_{\mathrm{Fe}})/2$ is assumed. $\rho\ell$ was calculated from the Drude theory.
\end{thebibliography}
\end{document}